\documentstyle[prl,aps,multicol,epsfig,tabularx]{revtex} 

\begin{document} 
 
\draft 
 
\title{Shocks in sand flowing in a silo}

\author{Azadeh Samadani$^{1}$, 
L. Mahadevan$^{2}$ \and A. Kudrolli$^{1}$}%

\address{$^{1}$ Department of
Physics, Clark University, Worcester, MA 01610, USA \\ $^{2}$
Department of Applied Mathematics and Theoretical Physics, Cambridge
University, CB3 9EW, UK}

\date{\today} \maketitle 


\begin{abstract}

We study the formation of shocks on the surface of a granular material
draining through an orifice at the bottom of a quasi-two dimensional
silo.  At high flow rates, the surface is observed to deviate strongly
from a smooth linear inclined profile giving way to a sharp
discontinuity in the height of the surface near the bottom of the
incline, the typical response of a choking flow such as encountered in
a hydraulic jump in a Newtonian fluid like water.  We present
experimental results that characterize the conditions for the
existence of such a jump, describe its structure and give an
explanation for its occurrence.

\end{abstract}

\begin{multicols}{2}

The flow of granular materials presents a multitude of experimental
and theoretical challenges in many-body physics, as manifested in the
unusual collective behavior of a large assembly of macroscopic grains
that interact with each other through collisions and friction.  The
phenomena of interest in quasi-static motion range from jamming
transitions, the formation and evolution of shear bands, and the
transmission of forces and acoustic signals in granular
packing~\cite{jaeger96,Rajchenbach00,liu98}. In the dynamical regime
associated with technological applications such as silo flows, other
questions of interest include the origin and maintenance of
fluctuations and their relation to mean flows in the bulk, and the
nucleation and evolution of avalanche like behavior in the vicinity of
the surface~\cite{nedderman92,BCRE,degennes99}.

A system of long-standing technological interest in this last
situation is associated with flow in a silo.  Building on an
understanding of simple properties such as the flow rate as a function
of the draining orifice size and the shape of the
silo~\cite{nedderman92}, recent work has focused on the details such
as the form of the velocity profile of the bulk flowing regions. 
In~\cite{samadani99} we studied the flow in the vicinity of the
orifice of a silo.  The predominant flow is restricted to a parabolic
region centered at the origin and may be interpreted in terms of
simple diffusion models~\cite{nedderman78,mullins74}.  Here, we
complement those studies by considering the dynamical evolution of the
form of the free-surface in a quasi two-dimensional silo flow.

\begin{figure}
\centerline{\epsfxsize=3.5in \epsfbox{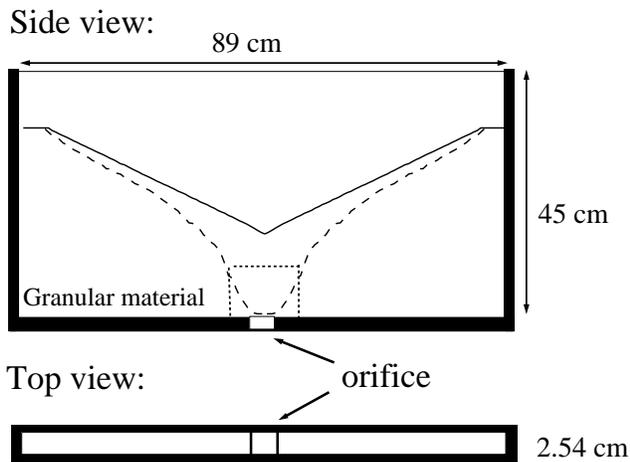}}
\caption{The experimental apparatus is a quasi-two dimensional silo 
with glass side walls. The dashed curve indicates an example of the boundary between rapidly flowing and static regions inside the silo. The dotted square box close to the orifice 
shows the approximate area corresponding to the images shown in Fig.~\ref{jump_1}(a) and Fig.~\ref{jump_1}(b).} 
\label{setup}
\end{figure}

Our system consists of a flat-bottomed silo that is
quasi-two-dimensional: 89 cm wide, 45 cm high and 2.54 cm deep, the
last being between 25-1000 times the typical grain size.  
The orifice itself has a square cross-section of width equal to that of 
the silo, with an attached valve to control the flow rate $Q$ 
(see Fig.~\ref{setup}). We use spherical glass grains with diameters 
$1$ mm, $500$ $\mu$m, $100$ $\mu$m, and $30$ $\mu$m to study the 
dependence of the phenomenon on grain size.  

The surface and the bulk of flow is visualized through the glass side 
walls of the silo using two types of CCD cameras.  The Kodak ES 1.0 
digital camera with a resolution of 1000 pixels $\times$ 1000 pixels 
with a maximum frame rate of 30 per second was used to measure the 
surface profile and to determine the location of the static and mobile 
regions inside the silo.  A second high speed Kodak camera with a 
maximum frame rate of 1000 per second and resolution of $256 \times 240$ 
pixels was used to measure the velocity of the particles.  By focusing 
in small regions where a particle corresponds to 3-4 pixels, we are able to
identify and track individual particles over consecutive frames with
an accuracy of $< 0.1$ pixels and thus determine the velocity.

When the orifice is opened, bulk flow is rapidly established and the
free surface initially subsides as an inverted Gaussian and then
quickly develops in to a V shape.  The angle of inclination of the
surface approximately corresponds to the angle of repose of the glass
beads.  For low flow rates, the free-surface has a constant slope, as
shown in Fig.~\ref{jump_1}(a).  However as the
flow rate is increased past a threshold by opening the valve, the
inclined free-surface is no longer linear but shows a region of abrupt
height transition right above the orifice, as shown in Fig.~\ref{jump_1}(b).

This unusual surface feature is similar to a hydraulic jump in a
Newtonian fluid observed when a stream of liquid encounters an
obstacle (such as observed when water impinges onto a flat plate).
Past work~\cite{morrison76,savage79,brennen83} has shown that
shock-like features similar to hydraulic jumps are observable in
granular flows in the presence of an externally introduced physical obstacle such
as a vertical plate in a channel flow. In these studies it was noted that the the ratio of inertial force to the gravitational force in the system given by the Froude number exceeds one, 
in analogy to the hydraulic jump observed in a fluid. Although, in
granular flow it is less obvious that this is important as in fluids where it corresponds to the flow velocity exceeding the wave speed for shallow depths. This fact was also noted earlier~\cite{savage79}.

\begin{figure}
     \centerline{\epsfysize=2.5in \epsfbox{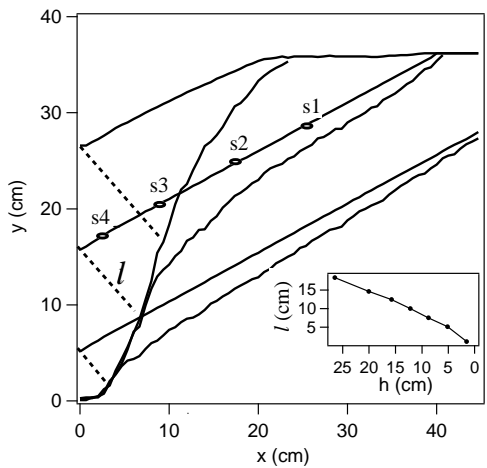}}
\caption{The surface and the boundary between flowing and static
regions at three different times.  The flow region is parabolic, and 
stay parabolic near the orifice during discharge. The orifice is centered 
at the origin $(0,0)$. The dashed lines show the size of
the intermediate region. The points s1, s2, s3, and s4 mark the locations 
where the velocity is measured and plotted in Fig.~\ref{difference}. 
$d =$ 1 mm.  Inset: The length $l$ that measures the size of the 
intermediate region as a function of height $h$ of the surface above 
the orifice is observed to decrease as the silo empties.  }
\label{schematic}
\end{figure}

In the case of the silo flow treated here, the jump is spontaneously generated by the flowing grains which creates its own ``obstacle'' by  choking right above the orifice. 
Moreover, unlike in a Newtonian fluid or in the channel flows of granular materials where
the entire fluid is in motion, here the grains exist in
different ``phases'' in different regions; away from the orifice and
the free-surface, they barely move at all, in the neighborhood of the
orifice they move slowly, near the free-surface they
flow like liquids, and veritably boil into a gas in the vicinity of the
free-surface shock [Fig.~\ref{jump_1}(c)]. 
To understand the formation of the free-surface shock it is important to understand the nature of the granular flow below the surface.  We
locate the surface and the boundary between the mobile and static
regions as the silo empties using the high resolution digital camera.
Two temporally separated images are subtracted  to determine the
boundary between static and mobile regions.  This
boundary along with the surface is shown for one-half of the silo in
Fig.~\ref{schematic} as the granular matter discharges.  

Since all the grains must eventually exit through the orifice, the vertical velocity
$v(x,y)$ is largest there, and the horizontal velocity $u(x,y)$ is
smallest.  However, as one moves away in the vertical direction $v$
gets smaller, while $u$ correspondingly becomes larger, and the extent
of the flowing region becomes larger.  As voids diffuse upwards from
the orifice, they explore a larger and larger distance in the
horizontal direction, and the bulk flow seems to be driven by
diffusive motion.  In fact, the flow in the silo close to the orifice
can be described remarkably well in terms of an invariant boundary
between flowing and non-flowing regions of the form $y \sim x^{2}/d$
with $d$ being the grain diameter~\cite{samadani99}, consistent with
the kinematic model for silo flows~\cite{nedderman78}. 

\begin{figure}
     \centerline{\epsfysize=2.5in \epsfbox{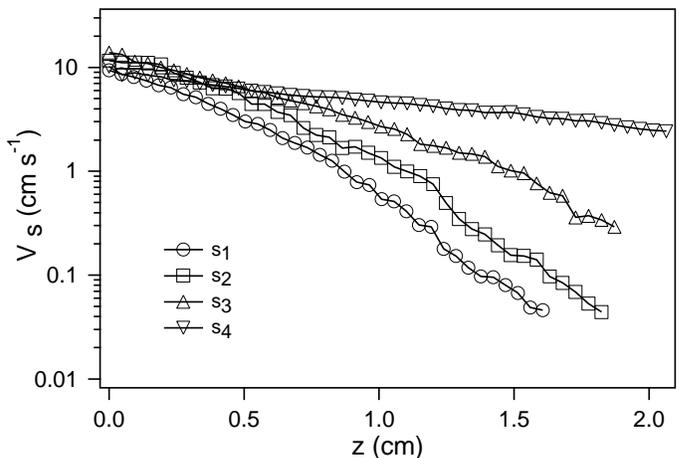}}
\caption{The velocity $v_s(s, z)$ as a function of depth $z$ at four
positions $s_1 - s_4$ along the surface (indicated in
Fig.~\ref{schematic}).  $d = $ 1 mm.  The $s$-axis is chosen parallel
to the surface and the $z$ axis is chosen perpendicular to the surface
and pointing inwards.  The velocities are observed to decay
approximately exponentially from the surface.  The length over which
the velocity decays $h_0$ is approximately 10-20 grain diameters.
Note also that the particles have reached an approximately constant
velocity $v_s(s, 0) = v_0$ along the surface.} \label{difference}
\end{figure}

The flow near the free surface outside this parabolic region
is confined to a shallow layer where particles move essentially parallel
to the surface.  However above the orifice, the flow near the free surface
changes from predominantly along the surface to a bulk flow towards
the orifice.  The size of this intermediate region is a function of
the height $h$ of the free surface above the orifice; this region is 
very large at early times, and gets progressively smaller. We parameterize this intermediate region using the distance $l$ from
the bottom of the surface to the point where the flow can be described
by the parabola (dashed line in Fig.~\ref{schematic}).  The inset to
Fig.~\ref{schematic} shows $l$ as a function of $h$.  Using the
high speed camera, we measured the velocity of particles
$v_{s}(s, z)$ in the neighborhood of four equidistant locations along
the free surface, with $s_{i}, i = 1-4$ indicated in Fig.~\ref{schematic} as a
function of the depth $z$ perpendicular to the surface. In Fig.~\ref{difference}, we plot $v_s(s_{i},z)$ vs. $z$ on a log-linear scale.
The velocity decays rapidly for the shallow flowing regions ($s_1, s_2$) and 
is described by $v_s(s, z) = v_0 \exp(-\alpha_{1} z/d - \alpha_{2}(z/d)^2)$
where $\alpha_{1}$ and $\alpha_{2}$ are dimensionless quantities that depend 
on the location of the surface. In the intermediate region ($s_3, s_4$), 
the velocity profile can be described by a single exponential 
$v_s(s, z) = v_0 \exp(-\alpha_{1} z/d)$.
Fig.~\ref{difference} also shows that the particles quickly reach a steady
velocity characterized by the balance between gravitational forcing
and inelastic collisions~\cite{footnote1}.

\begin{figure}
     \centerline{\epsfysize=2.5in \epsfbox{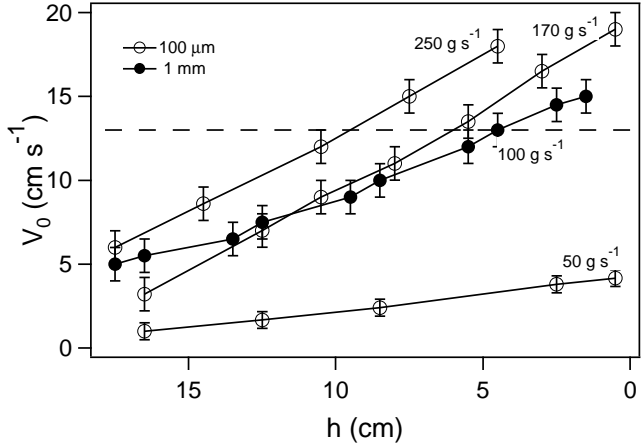}}
\caption{The velocity of grains $v_0$ at the surface as a function of
height above the orifice for $d = 100 \mu$m and $d = 1$ mm glass
beads.  The velocities are measured at a position just before the
location where the jump occurs ($s_4$ in Fig.~\ref{schematic}). 
The dashed line corresponds to the estimate of the critical velocity 
of 14 cm s$^{-1}$ for 100 micron particles. The critical velocity for 1 mm 
particles is approximately 44 cm s$^{-1}$.}
\label{velocity}
\end{figure}

We followed the free-surface as a function of time using the high
resolution camera (Fig.~\ref{surface}) while simultaneously measuring
$v_{0}$ (Fig.~\ref{velocity}) for the 100 $\mu$m glass beads.  In
Fig.~\ref{velocity}, we see that $v_0$ increases as the silo empties.
For low $Q$, the surface always remains V shaped throughout the draining
process.  For larger $Q$ the V shaped surface changes into a U shaped
surface eventually leading to the formation of a granular jump.
For 100 $\mu$m beads the shock appears at a height of about 10 cm for a $Q$ of 250 g
s$^{-1}$, and at a height of about 6 cm for a $Q$ of 170 g s$^{-1}$.  A
shock is not observed if the $Q$ is decreased further by using the
valve.  Thus the shock appears only when the surface velocity exceeds
a critical value and occurs at a higher height for higher $Q$ (dashed
line in Fig.~\ref{velocity}). 

\begin{figure}
     \centerline{\epsfysize=2.5in \epsfbox{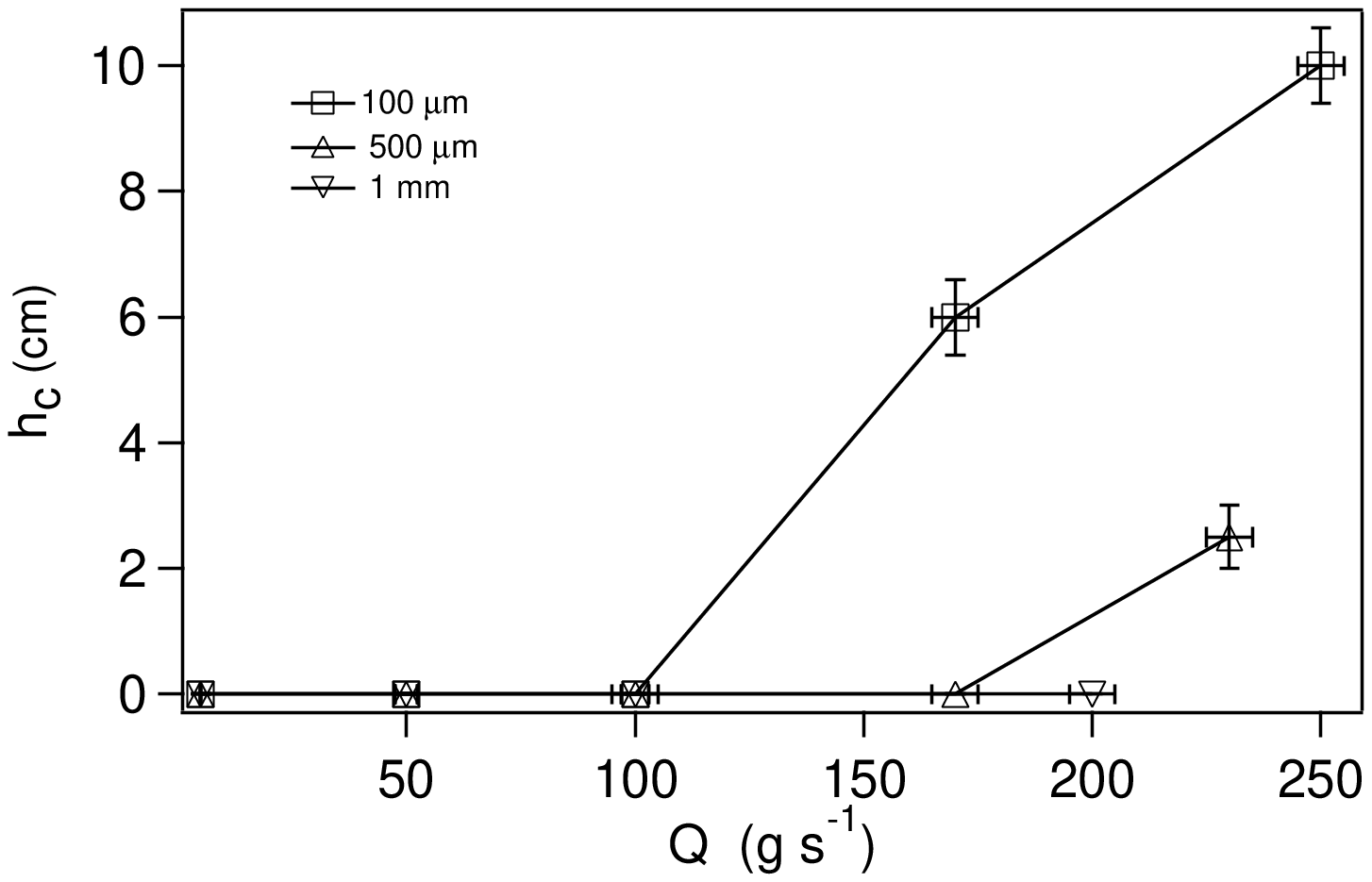}}
\caption{A plot of the critical height $h_c$ where the jump first appears as a function of the flow rate $Q$ for glass particles of various diameters $d$. A clear jump is not observed for  1 mm particles athough a small stagnant region is observed at the fastest flow rate.}
\label{phase}
\end{figure}

We also used beads of various sizes to study its effect on the appearance of the shock. It has been proposed that the flow rate $Q \propto (W - \alpha d)^{5/2}$, where $W$ is the width of the silo, and $\alpha$ is a constant between 0 and 1 depending on the geometry 
of the grain with size $d$~\cite{nedderman92}.  However, we observe that 
the size of particles has a greater influence on the flow rate than accounted 
for by this formula.  The maximum $Q$ for 30 $\mu$m, 100 $\mu$m, 500 $\mu$m 
and 1 mm beads was 260 g s$^{-1}$, 250 g s$^{-1}$, 230 g s$^{-1}$ and 200 g s$^{-1}$ respectively for the same 
(2.54 cm $\times$ 2.54 cm) orifice. In case of 30 $\mu$m beads,
the shock appears when the surface is approximately 30 cm above the
orifice as opposed to the 10 cm for 100 $\mu$m.  In case of 1 mm
beads, a clear shock does not develop although a stagnation zone at
the base of the incline surface is observed. We plot the critical height $h_c$ versus the flow rate for various $d$ in Fig.~\ref{phase} to characterize the transition of the surface from V-shape to that where a jump is observed. We note that the ratio of 
the drag force to weight of the particles $mg$, is less than $0.1\%$ 
for $1$ mm particles and less than $1\%$ for $100$ $\mu$m and therefore 
negligible. Therefore we believe that it is not the cause of the increase 
in height.  The drag force for $30$ $\mu$m particles is about $10\%$ and
may be important. The previous experiments~\cite{samadani99} for 2.54 cm 
and 1.27 cm wide silos, showed that the wall friction does not significantly
add to the drag force experienced by the particles as measured by the angle 
of repose and the shape of the parabolic flow region and therefore is negligible.

To rationalize these observations, we start by considering the increase in
$v_0$ which occurs for essentially geometric reasons.  The flow rate $Q$ is
proportional to the grain flux past the dashed lines shown in
Fig.~\ref{schematic}, and scales as $W \times \int v_s(s, z)dz$, where
$W$ is the width of the silo.  Using the observed exponential
dependence of $v_s(s, z)$, $Q \sim v_0 \times h_0$.  As the silo empties,
the bulk flow is confined to a smaller and narrower region (see
Fig.~\ref{schematic}). We have observed $Q$ by measuring the mass of the granular matter flowing out of the orifice in a fixed interval of time and find that $Q$ is constant to within 5\%~\cite{footnote2}. Since $Q$ is constant, $v_{0}$ must increase with time in the intermediate region whose
extent simultaneously decreases. Therefore the transition from
surface to bulk flow becomes more and more abrupt.  This increases the
frequency of inelastic collisions in the neighborhood of the valley
leading to a region of lower mean velocities (see
Fig.~\ref{jump_1}(c)).  The decrease in the horizontal
component of the velocity as the grains approach the center of the
silo leads to the conditions for a kinematic shock as faster grains
pile up behind (and atop) the slower grains.  Owing to the symmetry of
the flow from either side, the free surface responds by first
flattening along its axis of symmetry (see Fig.~\ref{surface}). As $v_0$
exceeds a critical value, an abrupt increase in the height of the free 
surface is the only recourse for the rapidly flowing surface grains which 
form a granular jump. 

In case of collisional granular flows, theoretical approaches have been developed first by~\cite{haff83} 
and~\cite{jenkins83} using mass-momemtum-energy-conservation to derive continuum equations. These approachs and subsequent improvements have been also reviewed by~\cite{campbell90}. However, applying these continuum approachs, especially in case of free boundary conditions as we have on one side of the flow and enduring contacts at the bottom of the surface flow, is extremely difficult. Therefore in the following we attempt scaling arguments to explain the phenomena.

To quantify the criterion for the formation of the
jump, we consider the analogous case of a hydraulic jump in a fluid
which occurs when the velocity of flow exceeds the propagation speed of a
surface perturbation. Dimensional arguments lead to an estimate of the speed
of propagation of surface perturbations as $v_p \sim \sqrt{g D}$, where 
$D$ corresponds to the depth of the flow~\cite{mahadevan99}.  In our case $D
\sim 20 \times d$.  Therefore $v_p \sim$ 14 cm s$^{-1}$ for 100 $\mu$m
beads, in qualitative agreement with the surface velocity at which the
shock is observed to occur. Finally, because $v_p$ increases with $d$, this is
also consistent with the occurrence of the shock at a much higher
height for 30 $\mu$m particles and the lack of a shock in case of the
larger 1 mm beads.

In conclusion, we have shown that a dynamic jamming event in silo
flows can lead to the formation of free surface shocks that are
similar to hydraulic jumps in simple fluids whose flow is obstructed.
An interesting feature of the granular jump is that the obstruction is
generated spontaneously by the flowing grains. We have given a kinematic
explanation for our observations, but in the absence of an elaborate
theory that describes granular flows, a quantitative
explanation remains an open question.

{\bf Acknowledgments} We thank D. Blair, E. Weeks, and W. Losert for fruitful
discussion.  This work was supported by the National Science
Foundation under grant number DMR-9983659 (Clark), the donors of the
Petroleum Research Fund (Clark) and the Office of Naval Research
through an NYI award (LM).  AK also thanks the Alfred P. Sloan
Foundation for its support; LM also thanks ENS-Paris and ESPCI for 
support via the
Chaire Condorcet and Chaire Paris Sciences.

\begin{figure}
\caption{(a) Image of granular matter draining from an orifice inside
a quasi-two dimensional silo as a function of time.  The surface is
linear in this case.  $Q \sim 50$ g s$^{-1}$, $d = 100 \mu$m.  The
images correspond to a 7 cm $\times \,7$ cm region close to the orifice.
(b) The surface is observed to show an increase in height similar to a
hydraulic jump when a rapidly flowing fluid in a channel encounters an
obstacle.  $Q = 200$ g s$^{-1}$.  (c) Closeup image ($\times$ 3) of the shock as it
is forming.  The duration of the exposure is 4 ms and therefore the
fast moving particles appear as streaks.  By comparison, the particles
inside the jump appear almost frozen.  A movie of the
jump can be viewed at http://physics.clarku.edu/$\sim$akudrolli/shock.html.} 
\label{jump_1}
\end{figure}

\begin{figure}
\caption{The surface profile of the granular matter draining from a
silo as a function of time showing the development of the jump.  ($Q
\sim 170$ g s$^{-1}$, $d =100 \mu$m).  The origin is located at the
center of the orifice. The first four surface profiles are separated by 5 second time intervals, to show the features of the jump, the last five are separated by 3 second time intervals. The dashed lines are guides to the eye to
compare the surface profile to that at low $Q$.  The surface flattens
and then a shock is observed to develop.}
\label{surface}
\end{figure}

\end{multicols}


\begin{thebibliography}{}

\bibitem{jaeger96} Jaeger, H., Nagel, S. R., and Behringer, R., Rev.  Mod. Phys. {\bf 68}, 1259(1996).

\bibitem{Rajchenbach00} Rajchenbach, J., Adv. Phys {\bf 49}, 229 (2000).

\bibitem{liu98} Liu, A. J., and Nagel, S. R., Nature  (London) {\bf 396}, 21 (1998).

\bibitem{nedderman92}Nedderman, R. M., Statics and kinematics of granular materials, Cambridge (1992).

\bibitem{BCRE} Bouchaud, J.-P., Cates, M., Raviprakash, J., and J. Edwards, J. Phys. I (France) {\bf 4}, 1383 (1994).

\bibitem{degennes99} de Gennes, P. G., Rev.  Mod. Phys. {\bf 71}, S374 (1999).

\bibitem{samadani99} Samadani, A., Pradhan, A., and Kudrolli, A., Phys. Rev. E {\bf 60}, 7203 (1999).

\bibitem{nedderman78} Nedderman, R. M., and Tuzun, U., Powder Tech. {\bf 22}, 243 (1979).

\bibitem{mullins74} Mullins, W. W., Powder Tech. {\bf 9}, 29 (1974).

\bibitem{morrison76} Morrison, H. L., and Richmond, O., J. Appl.  Mech. {\bf 43}, 49 (1976).

\bibitem{savage79} Savage, S. B., J. Fluid Mech. {\bf 92}, 53 (1979).

\bibitem{brennen83} Brennen, C. E., Sieck, K. and Paslaski, J., Powder Tech. {\bf 35}, 31 (1983).

\bibitem{footnote1}{An exponentially decaying velocity profile
as a function of depth has been reported recently in surface
flows~\cite{komatsu01,lemieux00} and in a Couette
geometry~\cite{losert00}.  However stretched-exponential profiles have
been also reported in a Couette geometry~\cite{mueth00}.  Our data
appears to be more consistent with a stretched exponential than a simple
exponential function in the shallow surface flow regime.}

\bibitem{footnote2}{We note that during the ultimate stages of draining, the sand empties like a Newtonian fluid under its own hydrostatic head; this could be a potential choking mechanism but in our experiments the shock typically forms much earlier, and
therefore must have its origins elsewhere.}

\bibitem{haff83}Haff, P. K., J. Fluid Mech. {\bf 134}, 401 (1983) .

\bibitem{jenkins83} Jenkins, J., and Savage, S., J. Fluid. Mech. {\bf 130}, 187 (1983).

\bibitem{campbell90} Campbell, C. S., Annu. Rev. Fluid Mech. {\bf 22}, 57 (1990).

\bibitem{mahadevan99} Mahadevan, L., and Pomeau, Y., Europhys.  Lett. {\bf 46}, 595 (1999).

\bibitem{komatsu01}  Komatsu,T. S., Inagaki, S., Nakagawa, N., and Nasuno, S., Phys.  Rev.  Lett. {\bf 86}, 1757 (2001).
\bibitem{lemieux00} Lemieux, P.-A., and Durian, D. J., Phys.  Rev.  Lett. {\bf 85}, 4273 (2000).

\bibitem{losert00} Losert, W., Bocquet, L., Lubensky, T. C., and Gollub, J. P., Phys.  Rev.  Lett. {\bf 85}, 1428 (2000).

\bibitem{mueth00} Mueth, D. M., Debregeas, G. F., Karczmar, G. S., Eng, P. J., Nagel, S. R., Jaeger, H. M., Nature (London) {\bf 406}, 385 (2000).

\end{thebibliography}
\end{document}